\documentstyle[twoside,fleqn,espcrc2,epsf]{article}

\def\centeron#1#2{{\setbox0=\hbox{#1}\setbox1=\hbox{#2}\ifdim
\wd1>\wd0\kern.5\wd1\kern-.5\wd0\fi
\copy0\kern-.5\wd0\kern-.5\wd1\copy1\ifdim\wd0>\wd1
\kern.5\wd0\kern-.5\wd1\fi}}
\def\centerover#1#2{\centeron{#1}{\setbox0=\hbox{#1}\setbox
1=\hbox{#2}\raise\ht0\hbox{\raise\dp1\hbox{\copy1}}}}
\def\centerunder#1#2{\centeron{#1}{\setbox0=\hbox{#1}\setbox
1=\hbox{#2}\lower\dp0\hbox{\lower\ht1\hbox{\copy1}}}}
\def\lsim{\;\centeron{\raise.35ex\hbox{$<$}}{\lower.65ex\hbox
{$\sim$}}\;}
\def\gsim{\;\centeron{\raise.35ex\hbox{$>$}}{\lower.65ex\hbox
{$\sim$}}\;}
\newcommand{\dslash}{\mathop{\not\!\! D}}

\title{Light hadron spectroscopy}

\author{D. K. Sinclair\address{High Energy Physics Division, Argonne National
        Laboratory,\\ 9700 South Cass Avenue, Argonne, Illinois, 60439, USA}%
\thanks{Work supported by the U.S. Department of Energy, Division of High
        Energy Physics, Contract W-31-109-ENG-38}%
\thanks{Plenary review talk presented at LATTICE'95, Melbourne, 11th -- 15th
        July, 1995}}

\begin{document}

\begin{abstract}
I review the progress that has been made in light hadron spectroscopy from
lattice QCD, since the LATTICE'94 conference in Bielefeld. 
\end{abstract}

\maketitle

\section{INTRODUCTION}

Calculation of the hadron spectrum has been a central goal of Lattice QCD for
20 years. A precision calculation of the hadron spectrum would be a validation
of QCD, and a validation of Lattice QCD as a calculational technique. It would
give us confidence when we apply lattice QCD to calculating other quantities.
In addition we would be able to clarify the nature of observed hadrons and
predict where others might be found. True precision hadron mass calculations
have been performed for heavy-quarkonium systems \cite{FNAL,NRQCD,JAPAN}, and
are beginning to appear for heavy-light systems \cite{HL}. Despite some
monumental efforts such as that by the GF-11 group \cite{GF-11}, the light
hadron spectrum remains considerably less well determined than one might hope
for. This is unfortunate, since this part of the hadron spectrum is the most
complex and least well understood. 

The reasons for this situation are well known. The light hadrons (with possible
exception of the glueballs) are large, their size being determined by the pion
Compton wavelength, so that large lattices are needed. Small lattice spacings
are needed to make contact with the continuum limit. The u and d quark masses
are small, making the Dirac operator ill-conditioned and expensive to invert.
There are a large number of excited states, which makes it difficult to isolate
the ground state contribution to (almost) any hadron propagator. At small quark
masses, hadron propagators become more noisy -- especially for large (time)
separations. Simulations with dynamical quarks suffer from critical slowing
down as the quark mass approaches zero. 

For these reasons, much of the work on light hadron spectroscopy has been
performed in the quenched approximation. Unfortunately, it is in just this
regime of light quarks that this approximation becomes the most suspect. Even
here, few measurements have used lattices of much more than 3.5fm in spatial
extent. Many calculations are restricted to u and d quark masses so large that
$m_\pi/m_\rho \gsim 0.5$, requiring considerable extrapolation to obtain 
physical results ($m_\pi/m_\rho \approx 0.18$).

Contributions since LATTICE'94 are listed in Tables~1--8. In each table, the
columns give $\beta$, lattice size, number of configurations and number of
source positions/configuration.

\begin{table}
\caption{QUENCHED STAGGERED}
\begin{tabular}{lllll}

&&&&\\
Gottlieb: &  5.7    & $8^3  \times 48$  &    600     &    6    \\
          &   "     & $12^3 \times 48$  &    400     &    6    \\
          &   "     & $16^3 \times 48$  &    400     &    6    \\
          &   "     & $20^3 \times 48$  &    200     &    6    \\
          &   "     & $24^3 \times 48$  &    200     &    6    \\
          &  5.85   & $12^3 \times 48$  &    200     &    6    \\
          &   "     & $20^3 \times 48$  &    200     &    6    \\
          &   "     & $24^3 \times 48$  &    200     &    6    \\
          &  6.15   & $32^3 \times 64$  &    115     &    8    \\
&&&&\\

COLUMBIA:  &   5.7    & $16^3 \times 40$  &            &           \\
&&&&\\

JLQCD:     &   5.85  & $16^3 \times 32$   &     60     &         \\
           &   5.93  & $20^3 \times 40$   &            &         \\
           &   6.0   & $24^3 \times 40$   &     50     &         \\
           &   6.2   & $32^3 \times 64$   &     40     &         \\
&&&&\\

Kim -- DKS: &  6.0   & $16^3 \times 64$   &    410     &    2    \\
            &   "    & $24^3 \times 64$   &    339     &    1    \\
            &   "    & $32^3 \times 64$   &    200     &    1    \\
&&&&\\

Kim -- Ohta &  6.5   & $48^3 \times 64$   &    50      &    1    \\

\end{tabular}
\end{table}

\begin{table}
\caption{QUENCHED WILSON}
\begin{tabular}{lllll}

&&&&\\
GERMAN:    &  6.0    & $16^3 \times 32$   &   1000     &         \\
&&&&\\

LANL:      &  6.0    & $32^3 \times 64$   &    150     &         \\
&&&&\\

JLQCD:     &  6.0    & $24^3 \times 64$   &    200     &         \\
           &  6.1    & $24^3 \times 64$   &  50--100   &         \\
           &  6.3    & $32^3 \times 80$   &     50     &         \\
&&&&\\

JLQCD:     &  6.0    & $24^3 \times 64$   &    1000    &         \\

&&&&\\    

SUZUKI:    &  5.7    & $16^3 \times 32$   &    20      &         \\

\end{tabular}
\end{table}

\begin{table}
\caption{QUENCHED CLOVERLEAF}
\begin{tabular}{lllll}

&&&&\\
UKQCD:     &  5.7    & $12^3 \times 24$   &    480     &          \\
           &  6.0    & $16^3 \times 48$   &    125     &          \\
           &  6.2    & $24^3 \times 48$   &    184     &          \\

\end{tabular}
\end{table}

\begin{table}
\caption{QUENCHED VALENCE}
\begin{tabular}{lllll}

&&&&\\
Liu -- Dong: &  6.0  & $16^3 \times 24$   &            &           \\

\end{tabular}
\end{table}

\begin{table}
\caption{IMPROVED-QUENCHED CLOVERLEAF}
\begin{tabular}{lllll}

&&&&\\
SCRI:      &  6.80    & $16^3 \times 32$  &            &           \\
           &  7.40    & $8^3  \times 16$  &            &           \\
           &  7.60    & $16^3 \times 32$  &            &           \\
           &  7.75    & $16^3 \times 32$  &            &           \\
           &  7.90    & $8^3  \times 32$  &            &           \\

\end{tabular}
\end{table}

\begin{table}
\caption{IMPROVED-QUENCHED IMPROVED}
\begin{tabular}{lllll}

&&&&\\
CORNELL:   &  6.8     & $5^3 \times 14$   &             &          \\
           &  7.1     & $6^3 \times 16$   &             &          \\

\end{tabular}
\end{table}

\begin{table}
\caption{FULL QCD STAGGERED}
\begin{tabular}{lllll}

&&&&\\
COLUMBIA:  &   5.7    & $16^3 \times 40$  &            &           \\

\end{tabular}
\end{table}

\begin{table}
\caption{FULL-QCD-STAGGERED CLOVERLEAF}
\begin{tabular}{lllll}

&&&&\\
SCRI:      &   5.6    & $16^3 \times 32$  &   100      &           \\

\end{tabular}
\end{table}

In addition, three groups, COLUMBIA, S.~Kim -- DKS and Gottlieb have
contributions on the chiral limit. Kilcup has presented calculations of the
connected and disconnected contributions to the $\eta'$ propagators using
staggered quarks. 

Before proceeding to summarize the above contributions, let me first make a few
general observations on directions pursued since LATTICE'94. For work prior to
this one should consult recent review articles such as those by Michael
\cite{Michael} and by Weingarten \cite{Weingarten}, and the collections of
parallel talks (and poster sessions) in the proceedings of LATTICE'94 and
earlier conferences in the series. 

In the past year the trend has been towards high statistics, rather than
larger lattices and/or smaller quark masses. Most of the work has been in
the quenched approximation. Many of these light hadron spectrum calculations
were performed as an integral part of other projects, such as matrix-element
calculations and/or heavy-light physics calculations.

This high statistics, sometimes combined with use of multiple sources/sinks to
isolate the ground state, has enabled the spectra to be measured with
sufficient precision to permit extrapolations in quark mass and lattice
spacing. Tests have been made as to how well one reproduces the spectrum of
hadrons containing strange valence quarks. 

Using Wilson quarks for spectrum calculations has the advantage that flavour
symmetry is exact, although this is at the expense of losing all chiral
symmetry. A perhaps more serious limitation is that Wilson quarks introduce
errors ${\cal O}(a)$. ($a$ is the lattice spacing.) Staggered quarks have the
advantage of keeping a vestige of chiral symmetry, at the expense of losing
flavour symmetry. While having the disadvantage that their lattice spacing is
effectively $2a$ they have the advantage that their errors are ${\cal O}(a^2)$.
An alternative solution is to use an improved Wilson-quark action, the
Sheikholeslami-Wohlert \cite{SW} or cloverleaf action, which has the flavour
symmetry of the Wilson action but errors ${\cal O}(a^2)$. 

Preliminary studies aimed at implementing the Lepage programme \cite{Lepage} of
using improved actions to allow use of larger lattice spacings and smaller (in
number of sites) lattices have been undertaken. 

Section~2 presents highlights of the above mentioned spectrum calculations.
In section~3 we discuss the chiral limit of the pion spectrum of quenched
lattice QCD. Section~4 gives a brief discussion of the work by Kilcup et al. on
the $\eta'$ mass, while section~5 is devoted to discussions and conclusions.

\section{HIGHLIGHTS OF SPECTRUM CALCULATIONS}

What follows are brief descriptions of recent contributions on light-hadron
spectroscopy, which have been summarised in Tables~1--8. Most of the figures in
this section were provided by the people mentioned, or prepared by me from data
provided by said persons. All work presented in this section should be
considered as preliminary. The light-hadron spectroscopy contributions of two
of the JLQCD projects was only a sideline to other calculations, and was not
available to me at the time of preparing this paper. I therefore refer you to
the parallel talks by Hashimoto \cite{Hashimoto} and Aoki \cite{Aoki}for
details (I thank M.~Okawa for bringing these works to my attention). 

\subsection{QUENCHED STAGGERED}

Gottlieb \cite{Gottlieb} has calculated the light hadron spectrum with
staggered quarks in the quenched approximation to lattice QCD. He used
``corner'' wall sources and point sinks. His lattice sizes ranged from $8^3
\times 48$ to $32^3 \times 64$. At $\beta=5.7$ and $\beta=5.85$ his quark
masses ranged from 0.01 to 0.16, while at $\beta=6.15$ they ranged from 0.005
to 0.08. 

\begin{figure}[htb]
\vspace{-1in}
\epsfxsize=3.5in
\epsfclipon
\epsffile{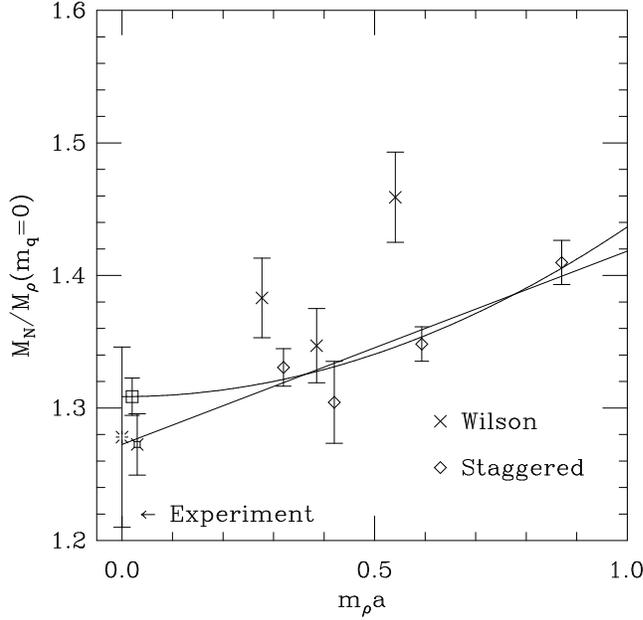}
\caption{$a$ dependence of $m_\pi/m_\rho$, showing linear and quadratic fits.}
\label{fig:gottlieb_a}
\end{figure}

High statistics on multiple lattice sizes has enabled him to analyse finite
size effects. For $m_\pi/m_\rho \gsim 0.5$, finite size effects appear to be
fairly small for spatial boxes of extent $>$~1.5--2fm. Use of several masses at
each beta has enabled him to extrapolate $m_N/m_\rho$ to $m_q=0$ (a linear
extrapolation appears adequate). Finally, use of several $\beta$'s and hence
$a$'s has permitted extrapolation to $a=0$. Both a linear and the accepted
quadratic extrapolations in $a$ give acceptable fits and mass ratios which are
too high by 5--8\%, which is normally considered acceptable considering the
systematic uncertainties of the ratios, fits and error estimates. (see
Figure~\ref{fig:gottlieb_a}). 

\begin{figure}[htb]
\vspace{-1in}
\epsfxsize=3.5in
\epsfclipon
\epsffile{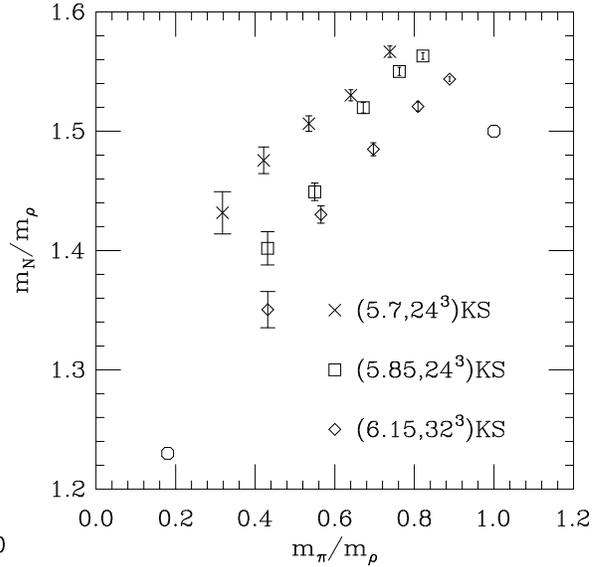}
\caption{Edinburgh plot -- quenched staggered.}
\label{fig:gottlieb_edinburgh}
\end{figure}

A plot of $m_N/m_\rho$ versus $m_\pi/m_\rho$ for these simulations is shown in
Figure~\ref{fig:gottlieb_edinburgh}.

Gottlieb \cite{GottliebJ} has also attempted to calculate the quantity $J$,
introduced by Lacock and Michael \cite{LM} as a consistency check on the meson
spectrum. ($J = m_{K^\ast} {dm_V \over dm_{PS}^2}$ evaluated at
$m_V/m_{PS}=m_{K^\ast}/m_K$.) His very preliminary results are: 0.56(1), for
$\beta=5.7$ ($24^3 \times 48$ lattice), 0.55(2) for $\beta=5.85$ ($20^3 \times
48$ lattice) and 0.49(1) ($24^3 \times 48$ lattice), and 0.45(1) for
$\beta=6.15$ ($32^3 \times 64$ lattice). A quadratic extrapolation to $a=0$
gives 0.43(1) compared with an experimental value of 0.48--0.50. He has also
studied the PCAC relationship for the pion mass (see Section~3). 

S.~Kim and D.~K.~Sinclair \cite{KS} have studied the light hadron spectrum with
quenched staggered quarks at $\beta=6.0$ (their previous simulations had been
at $\beta=6.5$ \cite{KSold}). They used wall sources and point sinks. Lattice
sizes of $16^3 \times 64$, $24^3 \times 64$, and $32^3 \times 64$ were used to
study finite size effects. Quark masses of $0.0025$, $0.005$, $0.01$ were used.

High statistics enabled us to study finite size effects. The $\pi$ showed no
significant finite size effects for box sizes $\gsim 24^3$; similarly for the
$\rho$ (at least for 2 heavier quark masses). Because of the difficulty in
finding a plateau in the nucleon effective mass on the $32^3 \times 64$
lattice, and because of the persistent difference between the two different
estimates (sources) for the nucleon mass, it remains unclear how significant
finite size effects are for the nucleon for these box sizes and quark masses. 

\begin{figure*}[htb]
\epsfxsize=6in
\epsffile{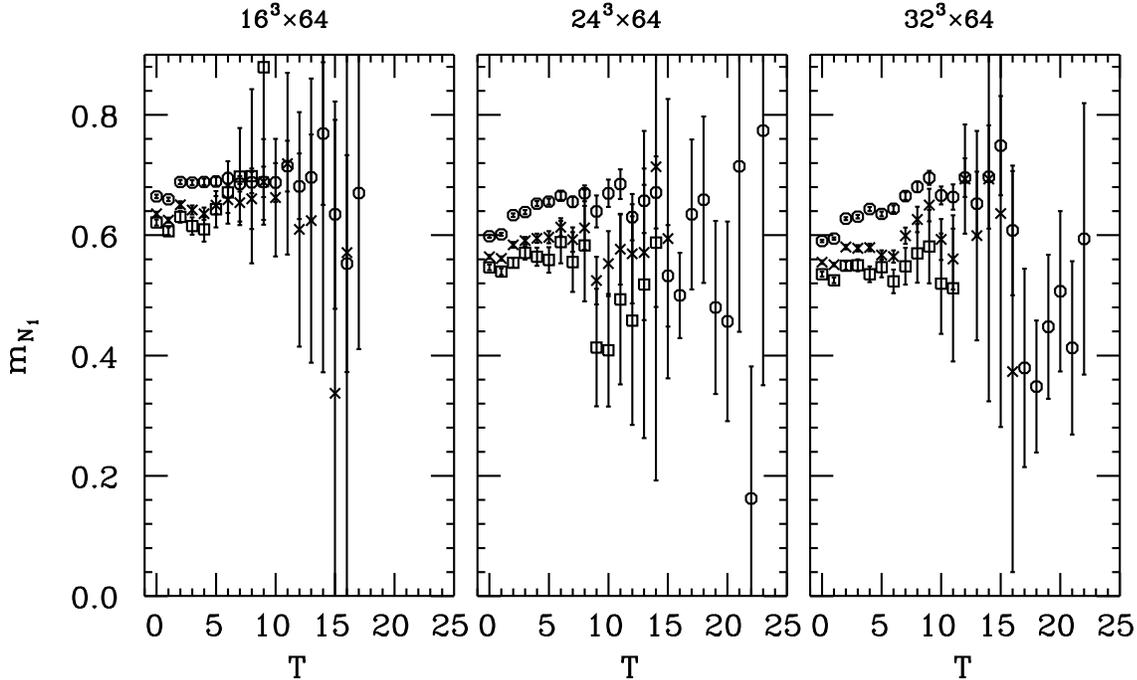}
\caption{Nucleon ($N_1$) effective mass plot -- quenched staggered.}
\label{fig:eff_mass_N1}
\end{figure*}

As one can see from the effective mass plots for the nucleon
(Figure~\ref{fig:eff_mass_N1}), at least for the largest lattice, it is unclear
whether a plateau is attained before the signal is swamped by noise, despite
the high statistics. We thus concluded that a wall source was a poor choice for
a $32^3 \times N_t$ lattice at $\beta=6.0$. 

\begin{figure}[htb]
\epsfxsize=3.0in
\epsffile{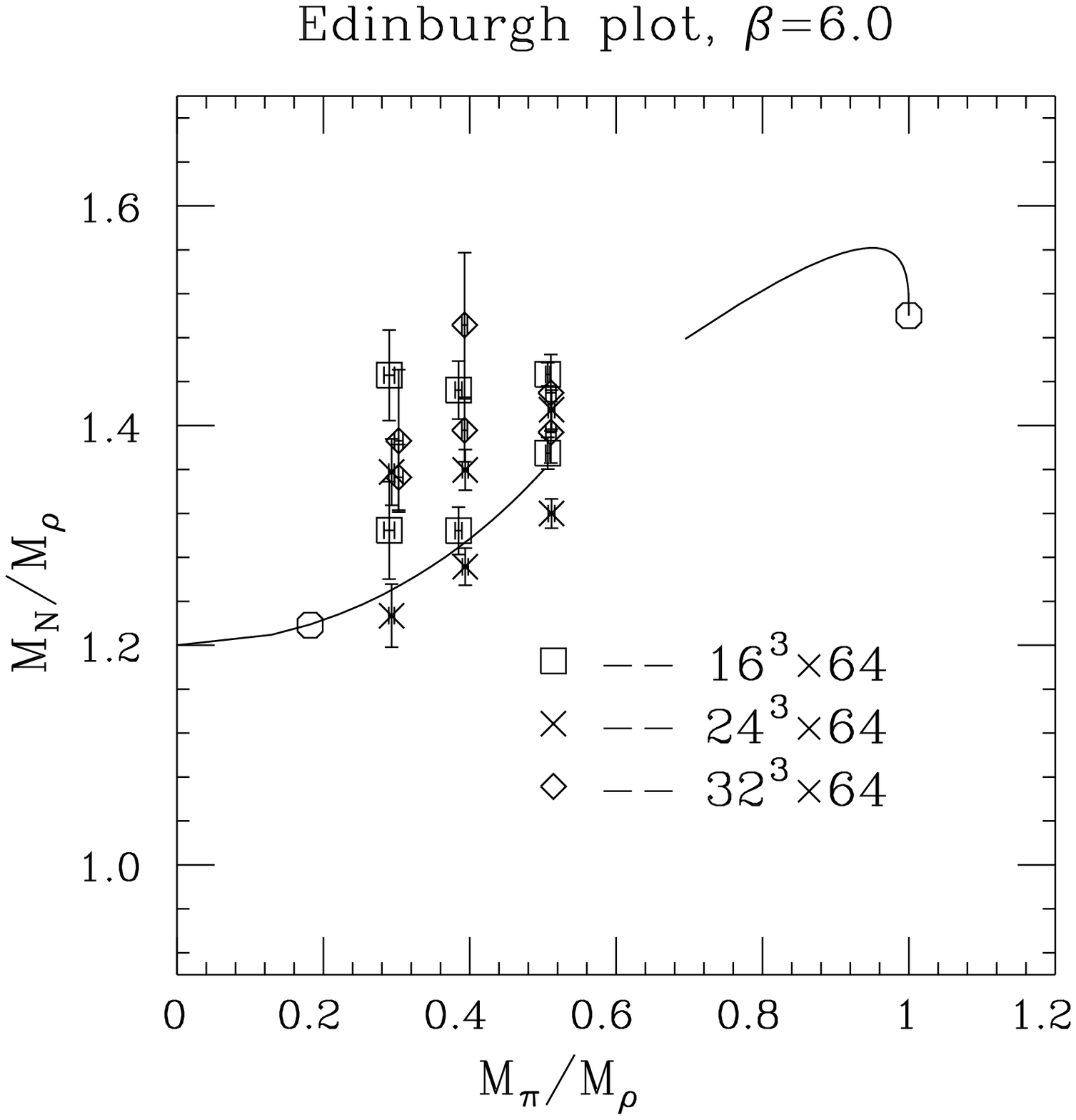}
\caption{Edinburgh plot -- quenched staggered.}
\label{fig:edinburgh}
\end{figure}

From the extrapolated $\rho$ mass, $a^{-1} = 1.87(2)$GeV at $\beta=6.0$. This
yields $a_{6.0}/a_{6.5} = 2.02(3)$ compared with scaling prediction
($\overline{MS}$) of 1.8976. Finally we include an Edinburgh plot 
(Figure~\ref{fig:edinburgh}). The 2 points for each mass/lattice size
come from the 2 different nucleon sources. 
A rough estimate of the J parameter, is $J \approx 5.6$, compared with the
experimental value of 4.8--5.0. Our results for the chiral behaviour of the
theory are presented in section 3. 

S.~Kim and S.~Ohta \cite{KO} are performing quenched simulations at $\beta=6.5$
on a $48^3 \times 64$ lattice. On their current sample of 50 lattices they have
calculated the light hadron spectrum with staggered quark masses 0.01, 0.005,
0.0025 and 0.00125 (in lattice units) enabling them to probe down to
$m_\pi/m_\rho < 0.3$. They use wall sources and point sinks. 

\begin{figure}[htb]
\epsfxsize=3in
\epsffile{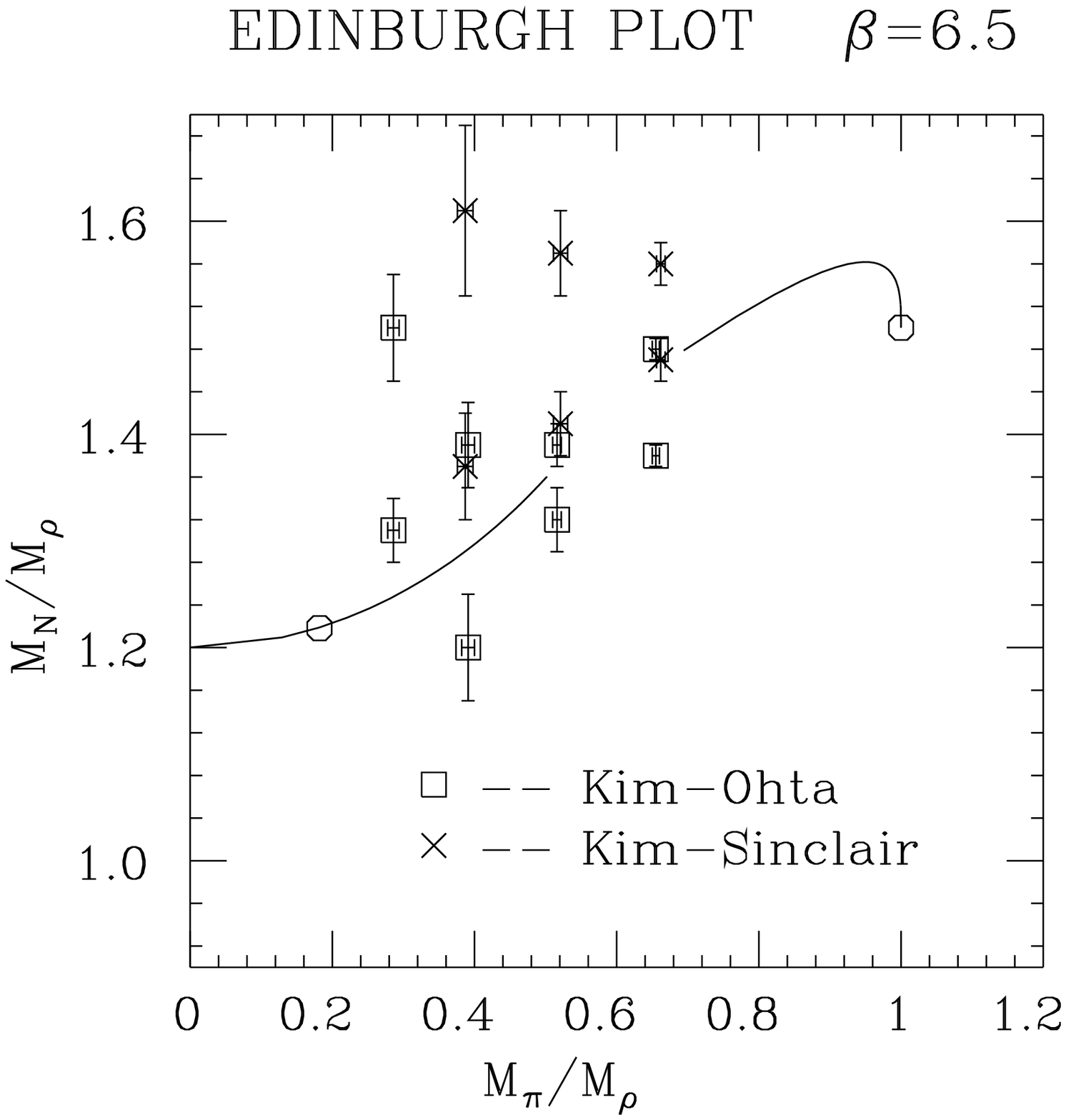}
\caption{Edinburgh plot -- quenched staggered.}
\label{fig:kim-ohta}
\end{figure}

When combined with the earlier work of Kim and Sinclair \cite{KSold}, this work
should enable them to study finite size effects at $\beta=6.5$, where scaling
is apparent and violations of flavour symmetry (indicated by the difference in
masses between Goldstone and non-Goldstone pions) are minimal. Their
preliminary results, summarized in the Edinburgh plot of
Figure~\ref{fig:kim-ohta} would seem to indicate considerable finite size
effects in the $m_N/m_\rho$ mass ratio. 
A word of caution is appropriate, since these are relatively low statistics,
preliminary results. $m_N/m_\rho$ decreases both because $m_N$ decreases
{\it and} because $m_\rho$ increases. Since the low statistics forces them to
fit propagators at relatively small time separations, at least some of the
difference could be due to not having yet reached the plateau in their
propagators. Only higher statistics can resolve this issue.

\subsection{QUENCHED WILSON}

The German collaboration (G.~Schierholz et al.) have performed quenched light
hadron spectroscopy with Wilson fermions as part of a nucleon structure function
calculation \cite{Schierholz}. They used a $16^3 \times 32$ lattice at
$\beta=6.0$, and used a Jacobi smeared source and sink. 

\begin{figure}[htb]
\vspace{-1in}
\epsfxsize=4.5in
\epsfclipon
\leavevmode
\epsffile{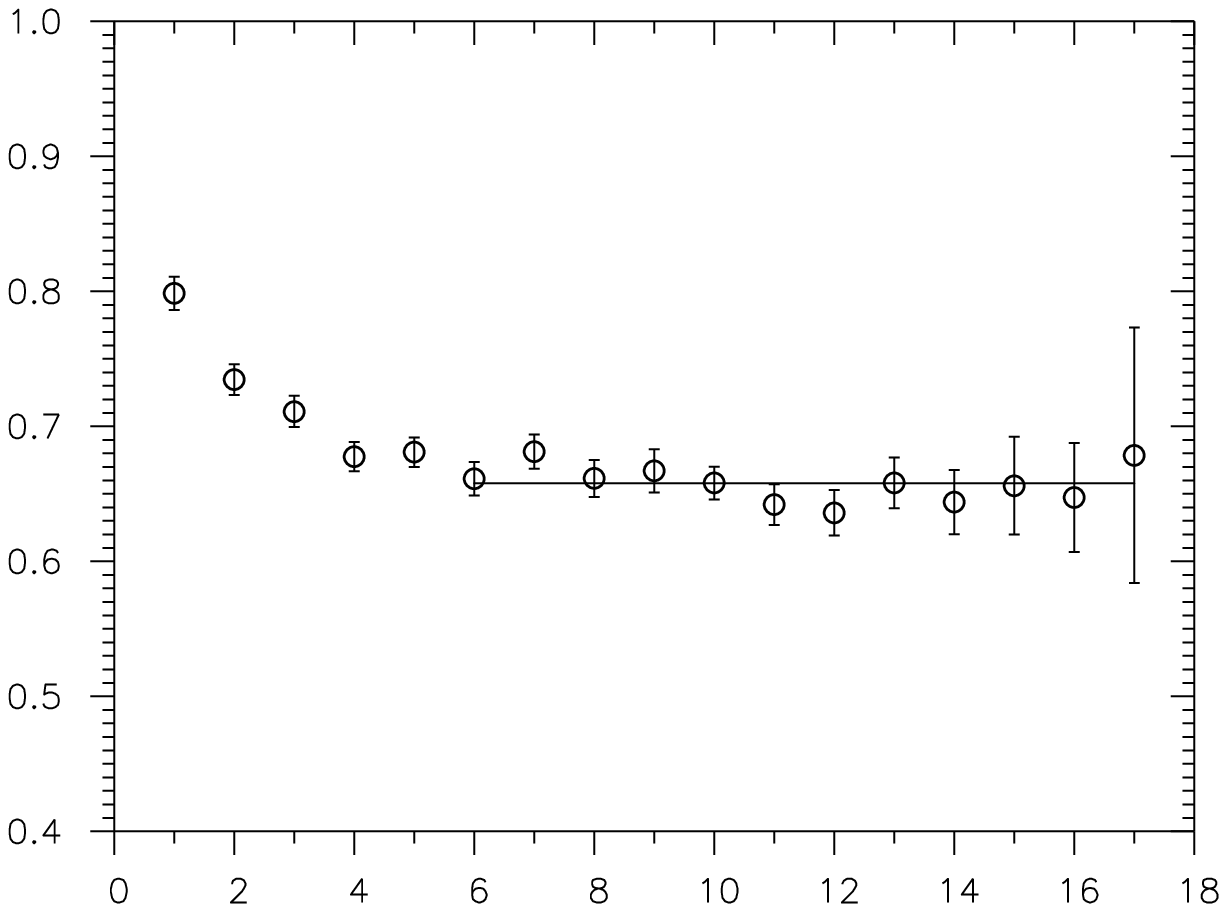}
\vspace{-2in}
\caption{Nucleon effective mass v's temporal separation.}
\label{fig:N_German}
\end{figure}

Using high statistics with Jacobi smearing (which appears to be a good choice),
they obtained well behaved hadron propagators with good evidence for plateaux 
(see Figure~\ref{fig:N_German} which shows the nucleon effective mass). 
They also use 2-component ``non-relativistic'' sources (projected with
$\frac{1}{2}(1+\gamma_4)$), which halves the number of required inversions.
The hadron mass values they obtain are given in Table~9.
\begin{table}
\begin{center}
\caption{The hadron masses in lattice units at $\beta=6.0$.}
\begin{tabular}{llll} \hline
 & \multicolumn{3}{c}{$\kappa$} \\ 
 & 0.1515 & 0.153 & 0.155  \\ \hline
$m_\pi$  & 0.504(1) & 0.422(2) & 0.297(2)  \\ 
$m_\rho$  & 0.570(2) & 0.507(2) & 0.422(2) \\ 
$m_N$  & 0.900(4) & 0.798(4) & 0.658(4) \\ \hline 
\end{tabular}
\end{center}
\end{table}

The LANL collaboration (communicated by R.~Gupta) have performed hadron
spectroscopy with quenched Wilson quarks on a $32^3 \times 64$ lattice at
$\beta = 6$ \cite{LANL}. They use four $\kappa$'s --- 0.153, 0.155, 0.1558,
0.1563 --- for u, d, s physics (they also do some charm physics). Hadron
propagators are obtained for wall source/point sink (WL), Wuppertal
source/point sink (SL) and Wuppertal source/Wuppertal sink (SS). Since SL and
SS effective masses approach the infinite separation limit from above, WL from
below, a linear combination was used to reduce the contributions from excited
states. 

\begin{figure}[htb]
\epsfxsize=3.0in
\epsfclipon
\epsffile{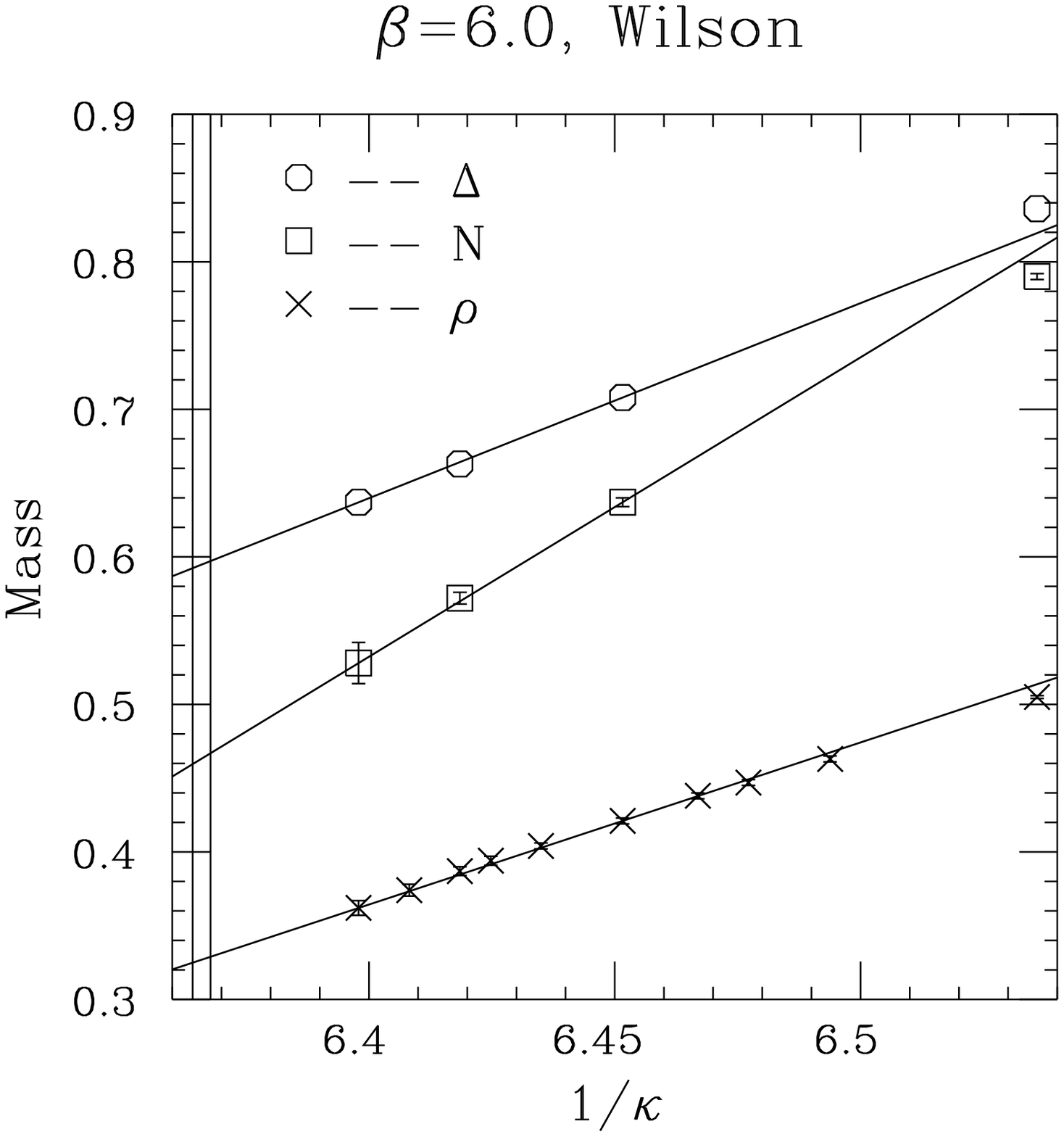}
\caption{Dependence of hadron masses on the quark mass}
\label{fig:gupta_m}
\end{figure}

\begin{figure}[htb]
\epsfxsize=3.0in
\epsffile{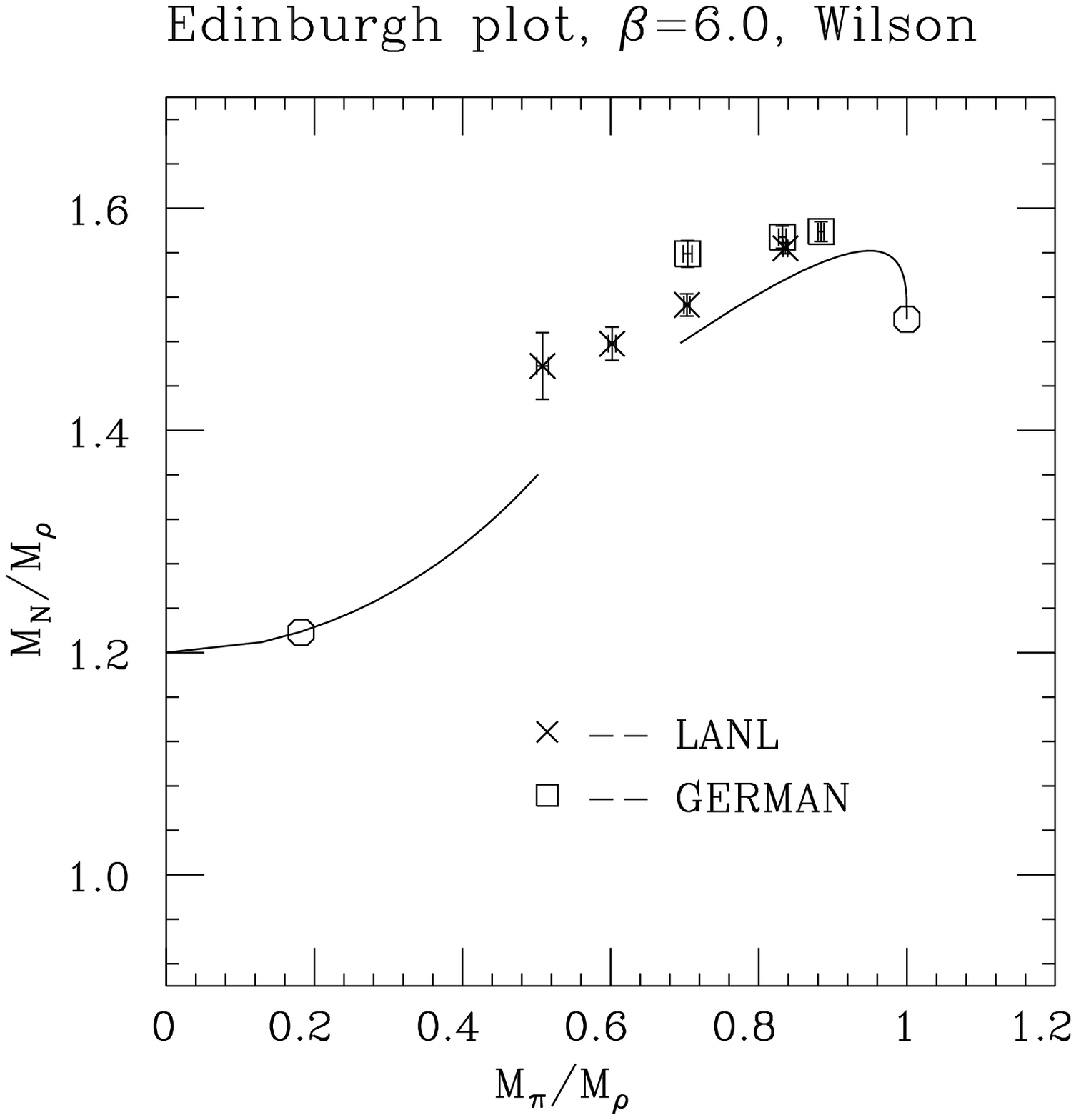}
\caption{Edinburgh plot -- quenched Wilson, LANL and GERMAN collaborations}
\label{fig:gupta_edinburgh}
\end{figure}

No significant finite size effects were observed between $24^3 \times N_t$ and
$32^3 \times 64$ lattices. Linear extrapolation to the correct pion mass gave
$m_N/m_\rho = 1.42(4)$, $m_\Delta/m_\rho = 1.80(5)$, (both 10 -- 15 \% too
high) and $a^{-1} = 2.319(44)$GeV (from $m_\rho$). Figure~\ref{fig:gupta_m}
shows the linear dependence of meson and baryon masses on the quark mass.
Figure~\ref{fig:gupta_edinburgh} is the Edinburgh plot from these simulations.

The values of $m_s$ required to give the correct values for $m_K/m_\rho$,
$m_{K^\ast}/m_\rho$, and $m_\phi/m_\rho$ differ by $\sim 20\%$. 
Even with the largest of these estimates for $m_s$, the baryon octet and
decuplet splittings are still as much as $\sim 30\%$ smaller than experiment.

Several forms for the lattice dispersion relation were tested for mesons. The 
best was found to be ${\rm sinh}^2(E/2) = {\rm sin}^2(p/2) + {\rm sinh}^2(m/2)$,
which worked well up to charmonium masses.

The JLQCD collaboration (presented by A.~Ukawa) have simulated quenched lattice
QCD on a $24^3 \times 64$ lattice at $\beta=6.0$ \cite{Ukawa}. They have
calculated hadron propagators and hence masses for 3 different $\kappa$ values
of Wilson quarks. Most of their preliminary results were for $\kappa=0.1550$
where $m_\pi/m_\rho \approx 0.70$. 

They have used their high statistics to examine the ``wiggles'' in the plateaux
of effective mass plots. Such behaviour is common. It is worrisome because
these fluctuations are often considerably larger than 1 standard deviation, and
thus unexpected. 

What Ukawa and his colleagues do is to calculate the complete covariance matrix
from the data. They then model the data with a propagator which is a 
2-exponential fit to the data, with gaussian fluctuations in the directions
of the eigenvectors of the true covariance matrix. The width of these gaussian
fluctuations is determined by the eigenvalues of this true covariance matrix.
Thus this model has the same average propagator and the same covariance matrix
as the actual data. Its effective mass plot for any finite sample shows 
deviations from the plateau by more than 1 standard deviation, as does the
real data. They thus conclude that the existence of such large fluctuations
is a property of the covariance matrix of the theory, rather than an indication
that no plateau exists or that their datasets are not independent. 

Suzuki and collaborators \cite{Suzuki} have studied quenched QCD with Wilson
quarks on a $16^3 \times 32$ lattice at $\beta=5.7$. Their study is aimed at
determining whether abelian monopoles, which have been suggested by 't~Hooft
\cite{'t Hooft}  to be responsible for confinement, are responsible for
generation of hadron masses in the chiral limit. 

Their procedure is as follows. After gauge fixing to the maximal abelian gauge,
they factorize the gauge fields into a diagonal abelian gauge field, and a
residual field. They then calculate field strengths corresponding to these
diagonal gauge fields. These abelian field strengths are separated into 
``smooth'', ``photonic'' contributions and Dirac string, monopole contributions.
Abelian gauge fields are reconstructed from each of these field strengths.

They have calculated Wilson quark and hence hadron propagators for each of the
full gauge field, the abelian gauge field, the monopole (Dirac string) gauge
field and the photonic gauge field. What they have found is that the $\pi$ and
$\rho$ masses for the full and abelian gauge configurations approach the same
values in the chiral limit. The results for the proton are unclear. More work
needs to be done before it will be known whether the masses for the monopole
gauge field yield the same results. The ``photon'' field case, as expected,
does not reproduce the correct spectrum in the chiral limit. 

\subsection{QUENCHED CLOVERLEAF}

I will start by discussing briefly the work of Lacock and Michael \cite{LM}, who
introduced the quantity $J = m_{K^\ast} {dm_V \over dm_{PS}^2}$ evaluated at
$m_V/m_{PS} = m_{K^\ast}/m_K \approx 1.8$. Experimentally J=0.49(2), if
they use the assumption that $m_{PS}^2 \propto (m_1+m_2)/2$ with $m_1$ and
$m_2$ the quark and antiquark masses in the pseudoscalar meson, $PS$ and
$m_V = c + d(m_1+m_2)$ with c and d constants. Ignoring for the moment any
isospin breaking, $m_\pi/m_\rho$ fixes the degenerate u and d quark masses,
while $m_K/m_\pi$ fixes the strange quark mass. Hence, on the lattice, $J$ is
determined. From the UKQCD data for the unimproved cloverleaf action they
found $J=0.37(3)(4)$, and similar values for Wilson data from other groups. 

The UKQCD collaboration (presented by H.~Shanahan \cite{Shanahan}) have studied
the light-light hadron spectrum with tadpole-improved cloverleaf quarks. They
ran at $\beta = 5.7$, $6.0$ and $6.2$ on lattices sizes up to $24^3 \times 48$.
Two $\kappa$ values were used at each $\beta$ ($\kappa = 0.14144$ and $0.14226$
at $\beta=6.2$). Point and Jacobi smeared and/or ``fuzzed'' sources and sinks
were used. 

Using multiple sources and sinks allowed stable 2-particle fits to propagators
and better determination of the ground state mass. They used smeared-smeared
(SS), smeared-local (SL) and local-local (LL) propagators in their fits.
Effective mass plots for each of these propagators for the vector states (from
their earlier runs with an unimproved cloverleaf action) are shown in
Figure~\ref{fig:UKQCD_eff_mass} \cite{LM}. Note that these various propagators
clearly have different couplings to the excited states, which allows stable
2-mass fits (shown in the figure). 
\begin{figure}[htb]
\vspace{-1in}
\epsfxsize=4.0in
\epsfclipon
\epsffile{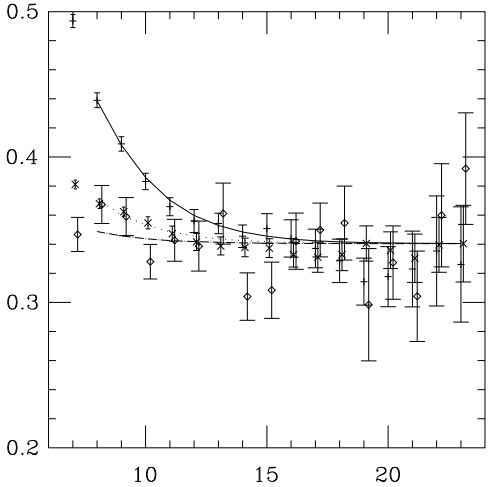}
\caption{Effective mass v's time separation for vector mesons, at 
         $\kappa=0.14266$ The symbols are LL(+), SL($\times$) and 
         SS($\Diamond$). The curves are the 2-state fits.}
\label{fig:UKQCD_eff_mass}
\end{figure}

J (defined above) was measured. The results were 0.368(5) ($\beta=5.7$),
0.330(19)(17) ($\beta=6.0$) and 0.363(60) ($\beta=6.2$), still below the 
experimental value.

$m_V^2-m_{PS}^2 \approx 0.55{\rm GeV}^2$ experimentally, from light mesons
through light-charm mesons. UKQCD's measured values of this quantity are in 
good agreement with this observation. Wilson quarks and unimproved cloverleaf
quarks are in considerably worse agreement with this observation. This is of
particular interest, since this splitting is due to the spin-orbit coupling.

\begin{figure}[htb]
\epsfxsize=3.0in
\epsfclipon
\epsffile{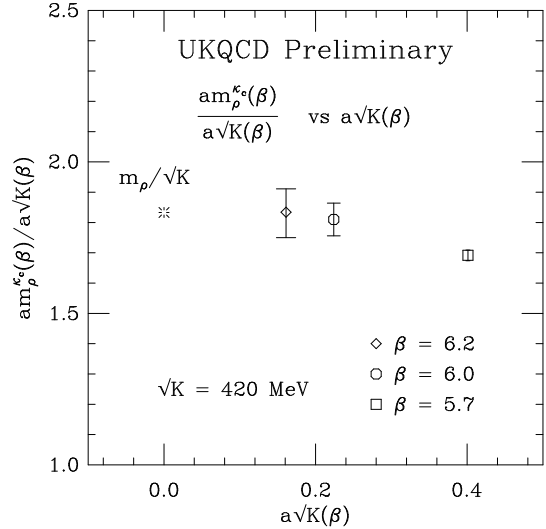}
\caption{$a$ dependence of $\rho$ mass. $K$ is the string tension.}
\label{fig:UKQCD_a}
\end{figure}

Figure~\ref{fig:UKQCD_a} shows the $a$ dependence of $m_\rho$ (obtained by
linear extrapolation in $m_q$). What one notices is that the points tend to
flatten out as one approaches $a=0$, which is what is expected, since
corrections are ${\cal O}(a^2)$ for the tadpole improved cloverleaf action.
This contrasts with the linear dependence on $a$ for the Wilson action. 

\subsection{QUENCHED VALENCE}

K.-F.~Liu and S.-J.~Dong have studied light hadron spectroscopy in the valence
quark approximation \cite{Liu}. They work on a $16^3 \times 24$ lattice at
$\beta=6$, with $\kappa=0.148$, $0.152$ and $0.154$. 

A word of explanation is needed since the quenched approximation itself is
often referred to as the valence quark approximation (even in this talk). Liu
and Dong's valence quarks are quenched Wilson quarks with all backward moving
contributions removed. Hence for meson propagators, exactly 1 quark and 1
antiquark pass through any time slice between source and sink. For baryons
exactly 3 quark lines pass through any timeslice. 

\begin{figure}[htb]
\epsfxsize=3.0in
\epsffile{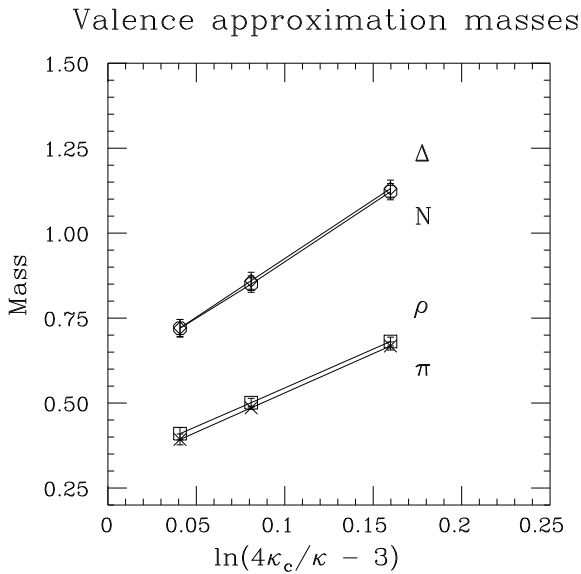}
\caption{$\Delta$,$N$,$\rho$ and $\pi$ masses as a function of quark mass in 
         the valence quark approximation.}
\label{fig:liu}
\end{figure}

Light hadron spectroscopy with valence quarks shows even better SU(6) symmetry
than one might expect. The $\Delta$ and nucleon masses are almost degenerate,
as are the $\rho$ and pion masses (in SU(6) $\Delta$ and N are in the same
multiplet, as are $\rho$ and $\pi$). This is shown in Figure~\ref{fig:liu}. 

They have also applied this approximation to the calculation of low energy 
matrix elements.

\subsection{IMPROVED-QUENCHED CLOVERLEAF}

The SCRI collaboration (presented by R.~Edwards \cite{Edwards}) are calculating
the light hadron spectrum in the quenched approximation. Gauge configurations
are being generated using the tadpole improved action of Lepage et al.
\cite{Lepage,LW}, with $\beta=6.8$--$7.9$ (note that these $\beta$ values
correspond to much smaller $\beta$'s for the Wilson action). Quark propagators
are being calculated using use tadpole improved cloverleaf action. Lattice
sizes are $8^3 \times 16$, $8^3 \times 32$ and $16^3 \times 32$. They are using
two gaussian smeared sources (and sinks?). Four $\kappa$'s corresponding to
fixed pion masses are being used. Fits to multiple correlation functions allow
2-exponential fits and better isolation of the ground state. 

\begin{figure}[htb]
\vspace{-1in}
\epsfxsize=2.5in
\epsffile{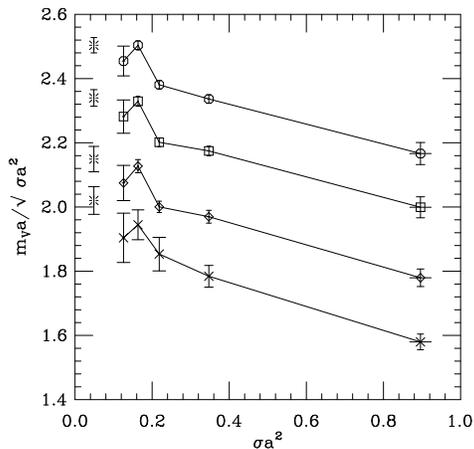}
\caption{$a$ dependence of the ($\rho$) mass. $\sigma$ is the string tension.}
\label{fig:edwards_a}
\end{figure}

This gauge action has errors ${\cal O}(a^4)$ (and ${\cal O}(\alpha_s^2 a^2)$),
compared with ${\cal O}(a^2)$ for the standard (Wilson) gauge action. The
cloverleaf quark action has errors ${\cal O}(a^2)$ (and ${\cal O}(\alpha_s^2
a)$). They are measuring the $a$ dependence of the spectrum. The meson spectrum
appears consistent with being linear in $a^2$ (the nucleon mass has yet to be
measured). The $a$ dependence of the $\rho$ mass is given in 
Figure~\ref{fig:edwards_a}.

The lattice spacings are relatively large (0.16 -- 0.41fm). (The chief advantage
of using the improved action is to allow this use of coarser lattices.) Of 
course, one really needs an ${\cal O}(a^4)$ quark action to make full use of
this fact.

$J = m_{K^\ast} {dm_V \over dm_{PS}^2}$, defined above, is measured and is
$\lsim 0.39$ for the parameters of these simulations (experiment $=0.48(2)$).
The potential and string tension are calculated as a scale for the hadron 
masses.

\subsection{IMPROVED-QUENCHED IMPROVED}

The Cornell group (presented by M.~Alford \cite{Alford}) has studied quenched
QCD with the tadpole improved action of Lepage et al.\cite{Lepage,LW}. The
quark and hence hadron propagators are generated on these configurations using
the D234 improved quark action. They work with $5^3 \times 14$ and $6^3 \times
16$ lattices at $\beta=6.8$ and $7.1$ respectively. 

The improved gauge action produces results which differ from the continuum
theory by terms ${\cal O}(a^4)$ (and ${\cal O}(\alpha_s^2 a^2)$. The D234
quark action is given by
\begin{eqnarray}
S_q & = &\bar{\psi}[-\dslash+m+{a^2 \over 6}D^{(3)} \nonumber \\
    &   &-{r a \over 2}(D^{(2)}+g F.\sigma)+s a^2 D^{(4)}]\psi
\end{eqnarray}
whose physics differs from the continuum by terms ${\cal O}(a^3)$ (and 
${\cal O}(\alpha_s a^2)$). This allows the use of much larger lattice spacings
than was possible with the Wilson gauge action, and Wilson, cloverleaf or
staggered quark action. In fact, with the chosen $\beta$'s the sizes of the
spatial boxes are $\sim 2$fm, which is larger than a $16^3$ spatial box at
$\beta=6.0$, with the Wilson gauge action.

The spectrum appears similar to that obtained from conventional simulations --
$m_N/m_\rho \sim $1.5 -- 1.6 for $m_\pi/m_\rho \sim $0.6 -- 0.65. The value of
the quantity J, defined above, is 0.38(2) at $\beta=6.8$ and 0.40(2) at
$\beta=7.1$
 
\subsection{FULL QCD STAGGERED}

The COLUMBIA group (presented by D.~Chen \cite{Chen}) have continued their
studies of the light hadron spectrum with 2 flavours of dynamical staggered
quarks. They ran on a $16^3 \times 40$ lattice, at $\beta = 5.7$, with
staggered quark mass $m_q = 0.01$. Two sources were employed for spectrum
calculations, a $16^3$ (wall) source and an $8^3$ (octant) source, with the 
gauge field fixed to coulomb gauge. 

Using 2 sources, with high statistics, enables 2-particle fits for the pion
propagator and 4-particle fits for other hadron propagators, enabling better
isolation of the ground state in each channel, and giving a preliminary
estimate for the mass of the first(?) excited state.

\begin{figure}[htb]
\epsfxsize=3.5in
\epsffile{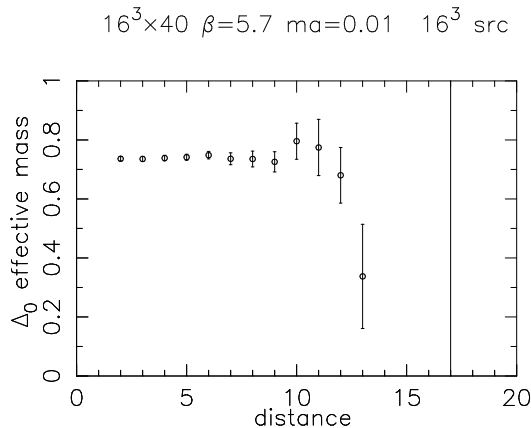}
\caption{$\Delta$ effective mass in Coulomb gauge}
\label{fig:columbia_Delta}
\end{figure}

For the $\Delta$ they tried Landau as well as Coulomb gauge. The mass from the
Landau gauge propagators was smaller than that from the Coulomb gauge
propagators. What is unclear is whether this is a true effect due to the
unphysical excitations in the Landau gauge, or just that the onset of the
plateau is delayed in the Landau gauge. Figure~\ref{fig:columbia_Delta} shows 
the $\Delta$ effective mass in coulomb gauge.

The chiral behaviour of the pion propagator as a function of valence quark 
mass has been calculated and will be discussed further in section~3.

\subsection{FULL-QCD-STAGGERED CLOVERLEAF}

The SCRI group (presented by J.~Sloan \cite{Sloan}) are calculating the light
hadron spectrum on the $16^3 \times 32$ HEMCGC configurations which were
generated with dynamical staggered quarks. The spectrum is being calculated
with tadpole improved cloverleaf quarks. $\beta=5.6$, and $m_q=0.01$ for the
dynamical quarks. Their spectrum calculations use wall and shell-smeared
sources. 

They employ fits to multiple propagators which enables better isolation of
the ground state. 

\begin{figure}[htb]
\epsfxsize=3in
\epsffile{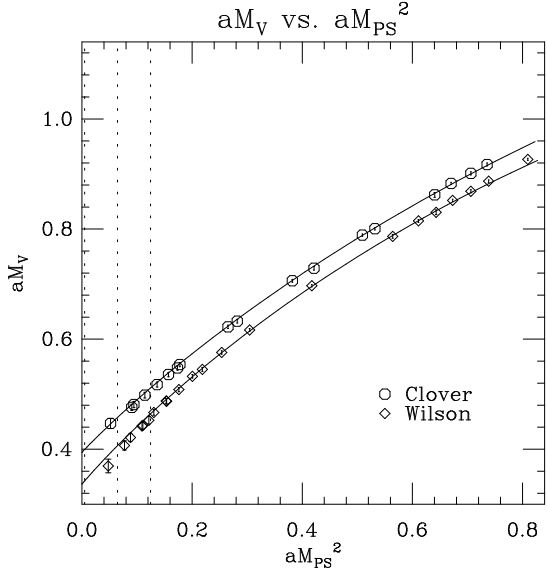}
\caption{Cubic fit of $m_V$ v's $m_{PS}^2$}
\label{fig:sloan_fits}
\end{figure}

Extrapolations of hadron masses to the chiral limit are being studied.
Vector meson masses are fit to forms which are polynomial in the pseudoscalar
meson mass $M_{PS}$. They have tried fits to polynomials up to fourth order,
but which do not include a linear term. Preferred fits were to the forms
\begin{equation}
M_V = C_0 + C_2 M_{PS}^2 + C_3 M_{PS}^3
\end{equation}
and
\begin{equation}
M_V = C_0 + C_2 M_{PS}^2 + C_4 M_{PS}^4
\end{equation}
The cubic fit above yields coefficients $C_0=0.393(4)$ and $C_2=1.10(3)$
for cloverleaf quarks. The data and these fits are given in 
Figure~\ref{fig:sloan_fits}. The fits for Wilson quarks are given for 
comparison.
From these chiral extrapolations of the $\rho$ mass they obtain $a^{-1} =
2.25(3)$GeV for Wilson quarks, $= 1.92(2)$GeV for cloverleaf. This is to be
compared with $1.80(2)$GeV obtained for staggered quarks. 

A linear approximation to $J$ defined above, yields 0.425(6)(10) for Wilson
quarks and 0.444(5)(26) for cloverleaf compared with the experimental value
0.499. Cloverleaf light vector meson masses are in reasonable agreement with
experimental values and closer than Wilson masses. Higher mass mesons are
predicted to be lighter than experimentally measured. 

\section{THE CHIRAL LIMIT --- CHIRAL LOGARITHMS v's
         FINITE SIZE v's FINITE LATTICE SPACING}

In quenched QCD, chiral perturbation theory predicts that
\begin{equation}
    m_\pi^2 = 2\mu m_q \left( m_\pi^2 \over \Lambda^2 \right)^{-\delta}
\end{equation}
rather than the usual
\begin{equation}
    m_\pi^2 = 2\mu m_q 
\end{equation}
\cite{Sharpe,BG} where $\delta$ is estimated to be $\approx 0.2$.

On the other hand finite lattice size effects would be expected to modify the
above relation (Equation~5) to
\begin{equation}
    m_\pi^2 = constant + 2\mu m_q
\end{equation}
where constant $\rightarrow 0$ as the spatial volume $V \rightarrow \infty$.
The COLUMBIA group point out that these two departures from the canonical
behaviour can be very difficult to distinguish \cite{COLUMBIA}. 

\begin{figure}[htb]
\epsfxsize=3in
\epsffile{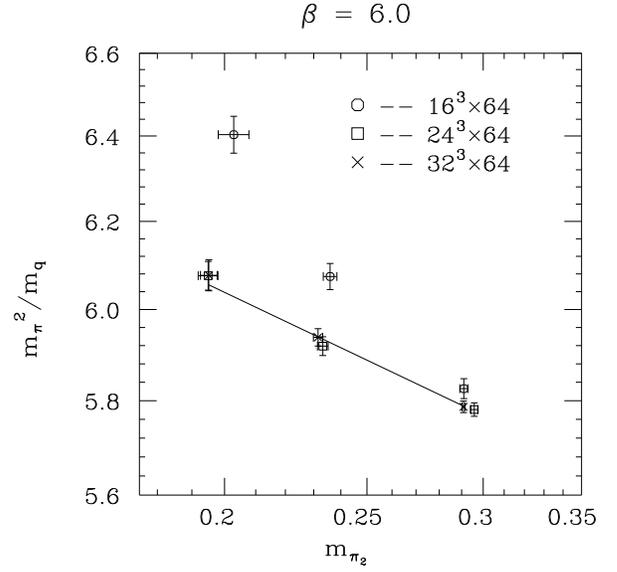}
\caption{$m_\pi^2/m_q$ v's $m^2_{\pi_2}$ at $\beta=6.0$ (quenched).}
\label{fig:chiral_us}
\end{figure}

In our $32^3 \times 64$ pion mass data we observed departures from the normal
relation above, which were well fit by the quenched chiral log formula
\cite{KSchiral}. Since there was good agreement between our $32^3 \times 64$
and $24^3 \times 64$ data, we rejected the interpretation of this as being a
finite size effect. We show this data in Figure~\ref{fig:chiral_us}. 

\begin{figure}[htb]
\epsfxsize=3in
\epsffile{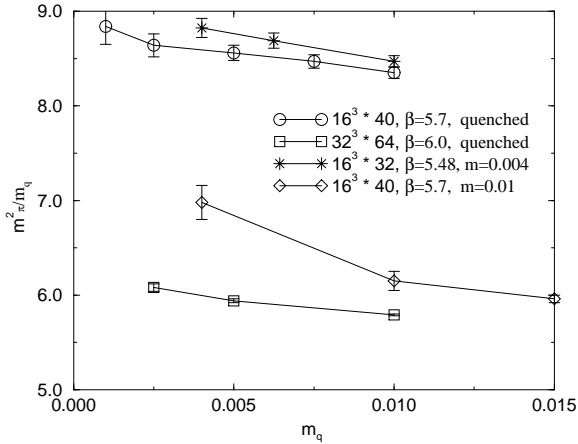}
\caption{$m_\pi^2/m_q$ v's $m_q$. Columbia results ($32^3 \times 64$ results
         are from S.~Kim and DKS).}
\label{fig:chiral_Columbia}
\end{figure}

The COLUMBIA group first observed similar departures from the canonical
relationship in full QCD at $\beta=5.48$ on a $16^3 \times 32$ lattice
and at $\beta=5.7$ on a $16^3 \times 40$ lattice when the dynamical quark mass
was held fixed (at $m_q=0.004$ and $0.01$ respectively) \cite{COLUMBIA}.
Although one would expect anomalous chiral behaviour in this case, it should
not be identical to the quenched case. They also performed quenched simulations
at $\beta=5.7$ on a $16^3 \times 40$ lattice and obtained similar results. 
For all their pion spectra they found good fits to the finite-size-modified
formula. They also managed to fit our results to the formula with a finite
intercept, although the fit had a much worse $\chi^2$. For their own
simulations, they were able to correlate these finite size effects in $m_\pi$
with those in $\langle\bar{\psi}\psi\rangle$. Figure~\ref{fig:chiral_Columbia} 
shows the variation of $m_\pi^2/m_q$ with quark mass.

\begin{figure}[htb]
\epsfxsize=3.5in
\epsfclipon
\epsffile{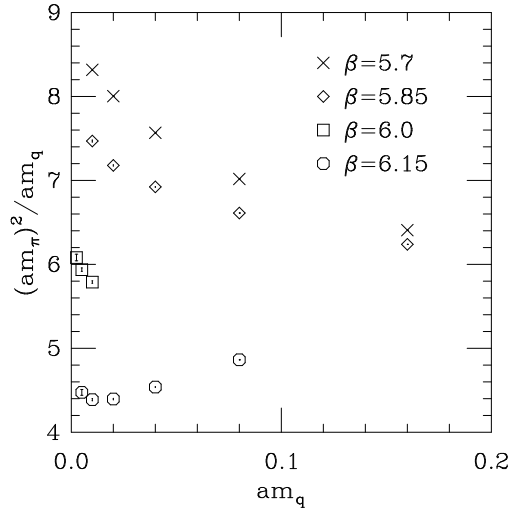}
\caption{$m_\pi^2/m_q$ v's $m_q$ for quenched QCD -- various $\beta$'s. (The
         $\beta=6.0$ data at higher masses is essentially flat.)}
\label{fig:chiral_Gottlieb}
\end{figure}

Gottlieb's quenched data at $\beta=5.7$, $5.85$ and $6.15$, combined with a
collage of data at $\beta=6.0$, indicates rather different variation of
$m_\pi^2/m_q$ with $m_q$ at different $\beta$'s suggesting that at least some
of the variation is a finite lattice spacing effect \cite{Gottlieb}. His data
at $\beta=5.7$ shows mass dependence out to masses for which finite size 
effects must be small 
(see Figure~\ref{fig:chiral_Gottlieb}).

\section{THE $\eta'$ MASS}

Kilcup et al. \cite{Kilcup} have calculated the $\eta'$ mass using a method
similar to that used by Kuramashi et al. \cite{Kuramashi}, but using staggered,
rather than Wilson quarks. They use $16^3 \times 32$ lattices at $\beta=6.0$
(quenched) and $\beta=5.7$ (full QCD -- Columbia group). The full QCD
configurations had $m_q=0.025$ and $0.01$. 

They calculate both the connected (1-loop) and disconnected (2-loop) 
contributions to the $\eta'$ propagator. The connected piece is calculated in
the standard fashion, but using a noisy source. The disconnected piece 
requires the inverse of the propagator at ``zero'' separation averaged over all 
sites of the source and sink time-slices (actually pairs of time slices with
staggered quarks). Since this is impractical, it is replaced by a stochastic
estimator, using 96 noise vectors per lattice.

The ratio of the 2-loop to 1-loop contributions to the propagator rises
linearly in time separation $t$ for quenched QCD. For full QCD it approaches a
constant as $t \rightarrow \infty$, exponentially. Such behaviour is evident in
their results. They estimate the $\eta'$ mass from the slope of the linear rise
in the quenched theory and from the mass scale in the exponential approach to
the asymptotic value for full QCD. Using the approximate value $a^{-1}=2$GeV,
they get $m_0=1050(170)$MeV for the quenched theory, and $m_0=730(250)$MeV for
the full theory, where 
\begin{equation}
m_{\eta'}^2 = m_0^2 + m_8^2
\end{equation}
and $m_8$ is the degenerate flavour octet ($\pi/K/\eta$) mass.

In addition they have studied the mass spectrum on cooled gauge field 
configurations and have confirmed that the $\eta'$ gets its mass from large
instantons. This agrees with the earlier work with Wilson fermions.

\section{SUMMARY AND CONCLUSIONS}

Meson spectroscopy is becoming precise enough to ask such questions as whether
the quenched approximation adequately describes mesons made of u, d, s quarks
and their antiquarks. Baryon spectroscopy is approaching this stage. However,
understanding of the systematic errors associated with choices of sources,
sinks, and choices of fits, is not yet good enough to preclude significant
disagreements between groups. (From Section~2 one would conclude that the
quenched approximation is less capable of reproducing the correct light hadron
spectrum than was claimed by the GF-11 group \cite{GF-11}.)

With Wilson quarks there does not appear to be any strange quark mass which
gives the correct light hadron spectrum in the quenched approximation, at the
lattice spacings currently used. Going to cloverleaf quarks does not appear to
improve the situation, and even using the Lepage improved gauge and quark
actions does not lead to a significant improvement. All these cases yield
values of J in the 0.36--0.40 range, whereas the experimental value is in the
range 0.48--0.50. Preliminary results for staggered quarks extrapolated to
$a=0$ give $J=0.43(1)$. The splittings in the baryon multiplets (Wilson quarks)
appear to be too small, for any reasonable value of the strange quark mass. 

With good statistics it appears possible to accurately extrapolate to the
``physical'' u and d quark masses. Such mass extrapolations, and the
interpolations used to include the effects of the s quark, are helped by the
fact that, at least for light quark physics, the assumption that hadron masses
depend on only the sum of valence quark masses appears valid. 

For full QCD, cloverleaf quarks are an improvement over Wilson quarks. The
value of J is better for cloverleaf than Wilson quarks, and for both Wilson and
cloverleaf quarks it is better than the quenched values. 

We have seen the first attempts at implementing the Lepage scheme for improved
actions. For more serious implementations this will require going to an 
${\cal O}(a^4)$ fermion action (to match the gauge action). This scheme shows
great promise as a calculational tool.

For the chiral limit of quenched lattice QCD with staggered quarks, it is
difficult to sort out chiral logarithms from finite size effects and finite
lattice spacing effects. If we have not seen chiral logarithms, the obvious
question is where are they, and what would it take to see them. Part of the
difficulty in observing chiral logarithms is the flavour symmetry breaking,
which gives the non-Goldstone pions and the disconnected part of the $\eta'$,
larger masses than the Goldstone pion. It would therefore be helpful to study
quenched QCD with Wilson/cloverleaf quarks, which have no flavour symmetry
breaking, down to such light pion masses. In conclusion, although the
difficulty in observing quenched chiral logarithms is bad news for theory, it
means that their effects are small, which is good news for phenomenology. 

Although we are still quite a way from doing precision calculations of the
light hadron spectrum with dynamical quarks, progress has been made,
particularly within the quenched approximation. Improved actions probably are
our best chance for full QCD. More work appears necessary to understand
systematics.

Note added in proof: Since the preparation of this manuscript we have received
a revised draft of the LANL preprint \cite{LANL}. In this, a more extensive
analysis of the baryon octet mass splittings reveals that these splittings
are not linear in the quark mass splittings, as had been previously assumed.
When this is properly taken into account, the baryon octet splittings are in
reasonable agreement with experiment. However, for the baryon decuplet, the
linear formula works well, and the disagreement with experiment remains. 

\section*{Acknowledgements}
I thank those who presented talks at the parallel sessions, and others who
have given me access to their work. In particular, I wish to thank those who
sent me figures and/or data which have been included in this talk. I apologize
to anyone who is unhappy with the description given above of his or her work,
and in particular for any errors that I might have introduced. In addition, I
would like to thank G.~T.~Bodwin for his help in solving the technical problems
associated with preparing this writeup.

\end{document}